\input{aipcheck}

\documentclass[
    ,final            % use final for the camera ready runs
  ]
  {aipproc}

\layoutstyle{6x9}

\begin{document}

\title{Foundation of  Hydrodynamics for Systems with Strong Interactions}


\author{Cheuk-Yin Wong }{
  address={Physics Division, Oak Ridge National Laboratory, Oak Ridge, TN 37830}
}

\begin{abstract}
For a dense and strongly interacting system, such as a nucleus or a
strongly-coupled quark-gluon plasma, the foundation of hydrodynamics
can be better found in the quantum description of constituents moving
in the strong mean fields generated by all other particles. Using the
result that the Schr\" odinger equation and the Klein-Gordon equation
can be written in hydrodynamical forms, we find that the probability
currents of the many-body system in the mean-field description obey a
hydrodynamical equation with stress tensors arising from many
contributions: quantum effects, mean-field interactions, and thermal
fluctuations.  The influence of various contributions to the
hydrodynamical motion is expected to vary with the temperature, as the
quantum and mean-field stress tensors playing more important roles at
low and moderate temperatures.
 
\end{abstract}

\maketitle

%%%%%%%%%%%%%%%%%%%%%%%%%%%%%%%%%%%%%%%%%%%%
%% MAINMATTER
%%%%%%%%%%%%%%%%%%%%%%%%%%%%%%%%%%%%%%%%%%%%

\section{Introduction}

The foundation of  hydrodynamics is
usually presented within the kinetic theory, in which particles are
described as a gas with weak interactions \cite{Hua63}. For 
a dense and strongly interacting system, such as a nucleus or a
strongly-coupled quark-gluon plasma, the foundation can be better
found in a quantum description of a system with constituents
moving in the strong mean fields generated by all other particles \cite{Won76,Won77}.   As the
time-dependent Schr\"odinger equation can be cast into a
hydrodynamical form \cite{Mad26,Won76},  the evolution
of the probability currents in the
mean-field potential provides support for the validity of the
liquid-drop model of a nucleus \cite{Won77}, as in Bohr and
Wheeler \cite{Boh38}, and the validity of the dynamical model of nuclear fission, as in Hill and Wheeler \cite{Hil53}.   The deviation of classical hydrodynamics from the quantum description is embodied in the presence of the quantum stress tensor $p_{ij}^{(q)}$ \cite{Won76,Won77}. Quantum shell effects manifest themselves
as nuclear shell effects superimposed on a smooth hydrodynamical
liquid-drop background, and they lead to the intrinsic deformation in
many nuclei \cite{Bra72}.

In particle and nuclear collisions at high energies,  a phenomenological relativistic hydrodynamical description of the process is a reasonable concept, as
pioneered by Landau \cite{Lan53} and supported by 
experimental findings \cite{Mur04,Ste05,Won08}.  In the related search
for the foundation of relativistic hydrodynamics using the Klein-Gordon
equation, we encounter the difficulty that the naive
probability density 
constructed from the wave function of the Klein-Gordon
equation may be negative.  The presence of a
negative probability density may appear to preclude its description in
relativistic hydrodynamics.  However, as pointed out by Feshbach and
Villars \cite{Fes58}, the negative probability density arises because the
Klein-Gordon equation describes both particle and antiparticle degrees of freedom.
Positive probability densities
can be formulated when we separate out the particle and antiparticle degrees of freedom 
to write  the Klein-Gordon equation  as a set of two coupled time-dependent Schr\" odinger
equations.  We shall follow and modify the formalism of Feshbach and
Villars \cite{Fes58} in order to search for the foundation for relativistic hydrodynamics in strongly-coupled systems.

\section{Non-relativistic Hydrodynamical Description of a Nucleus }

We begin by  briefly summarizing the non-relativistic hydrodynamical description of a nucleus with strongly interacting constituents. As the Schr\"odinger equation for a single-particle in
an external field can be cast into the form of a hydrodynamical
equation \cite{Won76}, a system of non-relativistic  single particles interacting in their own
mean fields can be described by non-relativistic hydrodynamics \cite{Won76,Won77}.  We can represent the
single-particle state wave function for a state $a$ in terms of
the amplitude function $\phi_a$ and the phase function $S_a$ as
\begin{eqnarray}
\label{decom}
\psi_a ({\bf r},t)=\phi_a({\bf r},t) \exp\{-i S_a({\bf r},t) \}.
\end{eqnarray}
In the time-dependent mean-field description we obtain the Euler
equation for the probability current  $nu^i$   \cite{Won77},
\begin{eqnarray}
\frac {\partial n u^i}{\partial t} + \sum_{j=1}^3
n u^i u^j =  -\frac{1}{m} \sum_{j=1}^3 
\frac{\partial}{\partial x_j} \left ( p_{ij}^{(q)}+p_{ij}^{(t)}+p_{ij}^{(v)}\right ),
\end{eqnarray}
where  $n=\sum_a n_a\phi_a^2$, $~~u^i=\sum_a n_a\phi^2 \nabla_i S_a /\sum_a n_a\phi_a^2$ for $i=1,2,3$, $n_a$ is the occupation number for the state $a$,  
\begin{eqnarray}
p_{ij}^{(q)}=-\frac{\hbar^2}{4m} \sum_a n_a \nabla^2 \phi_a^2 \delta_{ij}
+\frac{\hbar^2}{m} \sum_a n_a \nabla_i \phi_a \nabla_j \phi_a,
\end{eqnarray}
\begin{eqnarray}
p_{ij}^{(t)}=\frac{\hbar^2}{m} \sum_a n_a 
\phi_a^2 (\nabla_i S_a - m u^i) (\nabla_j S_a - m u^j),
\end{eqnarray}
\begin{eqnarray}
\label{non}
\frac{\partial}{\partial x^j} p_{ij}^{(v)}({\bf r},t)=
n({\bf r},t) \frac{\partial}{\partial x^j}
\int d^3 {\bf r}_2 n({\bf r}_2,t) v({\bf r},{\bf r}_2),
\end{eqnarray}
and $v({\bf r},{\bf r}_2)$ is the two-body interaction that generates the mean field.
The stress tensor due to the mean-field interaction can also be given as
\begin{eqnarray}
\frac{\partial}{\partial x^j} p_{ij}^{(v)}({\bf r},t)=
n({\bf r},t) \frac{\partial}{\partial x^j}
\left ( \frac{\partial (W^{(v)} n) }{\partial n} \right ),
\end{eqnarray}
where $W^{(v)}$ is the energy per particle arising from the mean-field interaction.  The mean-field stress tensor $ p_{ij}^{(v)}$ is given explicitly by 
\begin{eqnarray}
\label{pw}
 p_{ij}^{(v)}=
\left \{n \frac{\partial (W^{(v)} n) }{\partial n}
- W^{(v)} n \right \}\delta_{ij}.
\end{eqnarray}
Thus, the total pressure arises from many sources: (i) quantum stress
tensor $p_{ij}^{(q)}$ from quantum effects, (ii) the thermal stress
tensor $p_{ij}^{(t)}$ from the deviation of the individual velocity
fields from the local mean velocities, and the (iii) mean-field
stress tensor  $p_{ij}^{(v)}$ from the mutual interaction between fluid
elements.  For example, for a nucleus in which the nucleons interact
with the Skyrm interaction, we have \cite{Won77}
\begin{eqnarray}
 p_{ij}^{(q)}=
\frac{\hbar^2}{5 m}
\left ( \frac{6 \pi^2}{4} \right )^{2/3} n^{5/3} \delta_{ij} 
\end{eqnarray}
and
\begin{eqnarray}
 p_{ij}^{(v)}=\frac{3}{8} ( t_0 +\frac{1}{3} t_3 n ) n^2  \delta_{ij}, 
\end{eqnarray}
where for the Skyrm I nucleon-nucleon interaction, $t_0=-1057$
MeV/fm$^3$ is the two-body interaction strength, and $t_0=+14463$
MeV/fm$^6$ is the three-body interaction strength. 

A nucleus is a strongly-coupled system.  The quantum and 
mean-field stress tensors are the dominant component for the nuclear fluid at low and moderate temperatures.  The thermal stress tensor $p_{ij}^{(t)}$
can take on different values, depending on
the occupation numbers of the single-particle states that determines
the degree of thermal equilibrium of the system.  For a thermally
equilibrated system, the thermal stress tensor is
\begin{eqnarray}
 p_{ij}^{(t)}=
\frac{\hbar^2}{5 m}
\left ( \frac{6 \pi^2}{4} \right )^{2/3} n^{5/3} 
\left [ \frac{2mkT}{\hbar^2 m (6 \pi^2 n/4)^{2/3}} \right ]^2 \delta_{ij},
\end{eqnarray}
which is small for low and moderate temperatures.
We observe that when $|p_{ij}^{(q)}+p_{ij}^{(v)}| \gg p_{ij}^{(t)}$ for a strongly interacting system at low and moderate temperatures,  there can be situations in which the  system can behave quasi-hydrodynamically
 even though the state of the system has not yet reach thermal equilibrium, as is evidenced by the presence of nuclear collective vibrational and rotational states at low and moderate temperatures.  In this case, the hydrodynamical state is maintained essentially by the quantum stress tensor and the strong mean-field stress tensor, and not by the thermal stress tensor.

\section{Relativistic hydrodynamics and the Kelin-Gordon equation}

In the collision of high-energy particles and nuclei, the relativistic
hydrodynamics description was pioneered by Landau \cite{Lan53}.
Subsequent  considerations for the evolution
of quark-gluon matter in the work of Bjorken \cite{Bjo83}, Baym
$et~al.$ \cite{Bay83}, Ollitraut
\cite{Oll92,Oll08}, and many others \cite{hydro} lead to successful investigations of the
hydrodynamics of strongly interacting matter in extreme conditions in
relativistic heavy-ion collisions \cite{Mur04,Ste05,Won08,hydro}. 

In the search for the foundation of relativistic hydrodynamics
using the Klein-Gordon equation, we can overcome the apparent
difficulty of a  possible negative naive probability density
by re-writing the
Klein-Gordon equation as a set of coupled time-dependent Schr\"
odinger equations for particles and antiparticles, as in Feshbach and
Villars  \cite{Fes58}.   

 A general solution of the Klein-Gordon equation contains two
components that can be represented by a column vector,
\begin{eqnarray}
\label{Psi}
\Psi=
\left ( 
\begin{matrix} 
{ \chi_+\cr  
\chi_-  \cr} 
\end{matrix}
 \right ),
\end{eqnarray}
where $\chi_+$ and $\chi_-$ have positive norms, $|\chi_\pm|^2$, which
can be interpreted as the probability densities of particles and
antiparticles respectively.  
We find it necessary, however, to modify the formulation of Feshbach and Villars \cite{Fes58} so as to obtain a set of coupled equations that facilitate the re-writing of the Klein-Gordon equation in the hydrodynamical form.   We re-write the Klein-Gordon equation for 
the wave functions $\chi_+$
and $\chi_-$ 
as \cite{Won10}
\begin{eqnarray}
\label{sum}
(i\hbar\partial_{t}-e_{\pm}A^{0})\chi_{\pm}
\hspace{-0.35cm}
&=& \hspace{-0.35cm}
\frac{1}{2(E-e_a A^0)}\left \{\!\!(\frac{\hbar}{i}\nabla-e_{\pm}\vec{A})^{2}
\!\!+(m+S)^{2}
\!\!+[(E-e_a A^0)^{2}
\!\!-i\hbar \partial_t(E-e_a A^0)] \!\!\right \}\chi_{\pm}
\nonumber\\
&+& \hspace{-0.35cm}
\frac{1}{2(E-e_a A^0)}\left \{\!\!(\frac{\hbar}{i}\nabla-e_{\pm}\vec{A})^{2}
\!\!+(m+S)^{2}
\!\!-[(E-e_a A^0)^{2}
\!\!-i\hbar \partial_t(E-e_a A^0)]\!\!\right  \}\chi_{\mp}^*,
\nonumber\\
\end{eqnarray}
where $E$ is the positive expectation value of the operator
$i\hbar \partial_t$ for the single-particle state, $e_\pm=\pm e$, 
$e_a=\nu_a e$, and $\nu_a$ is the conserved particle number of the state.

\section{Klein-Gordon equation in hydrodynamical form}

To obtain the equation of continuity and the Euler equation from  Eq.\ (\ref{sum}), we write
the wave functions $\chi_\pm$ in terms of the amplitude and phase
functions as in Eq.\ (\ref{decom}),
\begin{eqnarray}
\chi_\pm({\bf r},t)=\phi_\pm ({\bf r},t) \exp\{iS_\pm({\bf r},t) \}.
\end{eqnarray}
We construct  $\chi_\pm^*$$\times$(\ref{sum})$-\chi_\pm$$\times$(\ref{sum})$^*$
and
obtain
\begin{eqnarray}
\label{con}
\partial_{t}[(E-e_a A^0) \phi_\pm^2]
+ \nabla\cdot [\phi_\pm^2(\nabla S_\pm -e_\pm \vec A) ]
=
X_\pm,
\end{eqnarray}
where 
\vspace{-1.0cm}
\begin{eqnarray}
2X_\pm &=&\{ \chi_\pm^*(\frac{\hbar}{i}\nabla-e_{\pm}\vec{A})^{2}\chi_\mp^*
- \chi_\pm(\frac{\hbar}{-i}\nabla-e_{\pm}\vec{A})^{2}\chi_\mp \}
\nonumber\\
& & +[(m+{\cal S})^2+(E-e_a A)^2] 
(\chi_\pm^*\chi_\mp^*-\chi_\pm \chi_\mp)
\nonumber\\
& & +
[i\hbar \partial_t(E-e_a A^0)] (\chi_\pm^*\chi_\mp^*+\chi_\pm\chi_\mp).
\end{eqnarray}
As the quantities $X_+$ and $X_-$ are not separately a full divergence,
the total number of particles and antiparticles
are not separately conserved.  However, the difference of the particle number and
antiparticle number of the two components satisfies the equation
\begin{eqnarray}
\partial_{t}[(E-e_a A^0) (\phi_+^2-\phi_-^2)]
+ \nabla\cdot [\phi_+^2(\nabla S_+ -e_+\vec A)] 
- \nabla\cdot [\phi_-^2(\nabla S_- -e_- \vec A)] 
=
X_+-X_-.
\end{eqnarray}
It can be shown that the difference $(X_+ -X_-)$ is a complete divergence. 
Therefore, the quantity 
\begin{eqnarray}
\label{eq38}
\nu_a=\int d{\bf r} \frac {E-e_a A^0}{m} (\phi_+^2-\phi_-^2)
=\int d{\bf r} \frac {E-e_a A^0}{m} (|\chi_+|^2-|\chi_-|^2)
\end{eqnarray}
is a conserved quantity.

Hydrodynamical description will be appropriate after the active pair
production stage has passed and the expansion of the system is driven
by a slowly varying mean field.  It is this type
of motion, with the suppression of pair production, for which we wish to provide a hydrodynamical description.
 Under such a circumstance, one can speak of a single-particle
system with a definite particle number $\nu_a$ which can take on
the value $\nu_a=1$ for a particle and $\nu_a=-1$ for an antiparticle. 
From Eq.\ (\ref{sum}), the equation for the phase function
$S_\pm$ for this simplified case is
 \begin{eqnarray}
(\partial_t S_\pm+e_\pm A^0 ) 
&=&
\frac{1}{2(E-e_a A^0)} \biggl  \{
[(\nabla^2 \phi_\pm)/\phi_\pm-(\nabla S_\pm
-e_\pm \vec A )^2 ]
\nonumber\\
& &
-(m+{\cal S})^2-(E-e_a A^0)^2  \biggl  \}.
\end{eqnarray}
For this case with suppressed pair production, $e_a=\nu_a e = e_\pm$.
We take the gradient $\nabla_i$ of the above for $i=1,2,3$, and
multiply by $\phi_\pm^2(E-e_\pm A^0)$.  
Using the equation of continuity for this simplifying case with
the suppression of pair production, we obtain
\begin{eqnarray}
& &
\partial_t\left [ \frac{(E-e_\pm A^0)\phi_\pm^2(\nabla_i S_\pm-e_\pm A^i)}{m+{\cal S}}\right ] +\sum_{j=1}^3 \nabla_j
\left [\frac {\phi_\pm^2 (\nabla_j S_\pm-e_\pm  A^j )  (\nabla_i S_\pm-e_\pm A^i ) }{m+{\cal S}} \right ]
\nonumber\\
&=&\hspace{-0.3cm}
-\frac{m}{m+{\cal S}}\sum_{j=1}^3 \nabla_j  p_{ij}^{(q)}
 -\phi_\pm^2 \nabla_i {\cal S}
\nonumber\\
& &\hspace{-0.3cm}
+\frac{1}{m+{\cal S}} \biggl \{ 
- (E-e_\pm A^0)\phi_\pm^2e_\pm F^{0i}
-\sum_{j=1}^3 \phi_\pm^2 (\nabla_j S_\pm-e_\pm  A^j )e_\pm F^{ij}
\biggr \}
\nonumber\\ 
& &\hspace{-0.3cm}
+\frac{e_\pm \nabla_i  A^0\phi_\pm^2}{2(E-e_\pm A^0)(m+{\cal S})} 
\biggl  \{\!\!\!
(\nabla^2 \phi_\pm)/\phi_\pm-(\nabla S_\pm-e_\pm \vec A )^2 ]
-(m+{\cal S})^2+(E-e_\pm A^0)^2\!\!\!\biggr \}
\nonumber\\ &  &\hspace{-0.3cm}
-(E-e_\pm A^0)\phi_\pm^2(\nabla_i S_\pm-e_\pm A^i) 
\frac{\partial_t {\cal S}} {(m+{\cal S})^2}
\nonumber\\ 
& &\hspace{-0.3cm}
-\sum_{j=1}^3 
\left [\frac {\phi_\pm^2 (\nabla_j S_\pm-e_\pm  A^j )  (\nabla_i S_\pm-e_\pm A^i ) }{m+{\cal S}} \right ]
\frac{\nabla_j {\cal S}} {(m+{\cal S})^2}. 
\end{eqnarray}
We can identity the fluid energy density $\epsilon_\pm$ as 
\begin{eqnarray}
\epsilon_\pm=(m+{\cal S})\phi_\pm^2,
\end{eqnarray}
as it corresponds to the fluid energy density for a fluid element at rest.
The fluid is characterized by a relativistic 4-velocity $u^\mu$.  We
can identify
\begin{eqnarray}
u_\pm^0&=&\frac{E-e_\pm A^0}{m+{\cal S}},
~{\rm and} ~~u_\pm^i=\frac{\nabla_i S_\pm-e_\pm A^i}{m+{\cal S}}, {\rm ~~~~for ~} i=1,2,3,
\end{eqnarray}
which obeys $(u_\pm^0)^2 - ({\vec u}_\pm)^2=1$ in the absence of the
quantum effects.  We can write the equation of motion for the
probability currents in terms of the hydrodynamical equation
\begin{eqnarray}
\label{hyd}
  & &\partial_t ( \epsilon_\pm u_\pm^0 u_\pm^i)
  + \sum_{j=1}^3 \nabla_j  \epsilon_\pm u_\pm^i u_\pm^j
  +\frac{m}{m+{\cal S}}   \sum_{j=1}^3 \nabla_j p_{\pm ij}^{(q)}
  \nonumber\\
  & =&
  -\phi_\pm^2 \nabla_i {\cal S}
  +\frac{1}{m+{\cal S}} \biggl \{ -
  (E-e_\pm A^0)\phi_\pm^2e_\pm F^{0i}
  -\sum_{j=1}^3 \phi_\pm^2 (\nabla_j S_\pm-e_\pm  A^j )e_\pm F^{ij}
  \biggr \}
  \nonumber\\ 
  & &
  +\frac{e_\pm \nabla_i  A^0\phi_\pm^2}{2(E-e_\pm A^0)(m+{\cal S})} 
  \biggl  \{\!\!\!
  (\nabla^2 \phi_\pm)/\phi_\pm-(\nabla S_\pm-e_\pm \vec A )^2 ]
  -(m+{\cal S})^2+(E-e_\pm A^0)^2\!\!\!\biggr \}
  \nonumber\\ &  &
  -(E-e_\pm A^0)\phi_\pm^2(\nabla_i S_\pm-e_\pm A^i) 
  \frac{\partial_t {\cal S}} {(m+{\cal S})^2}
  \nonumber\\ 
  & & 
 -\sum_{j=1}^3 
  \left [\frac {\phi_\pm^2 (\nabla_j S_\pm-e_\pm  A^j )  (\nabla_i S_\pm-e_\pm A^i ) }{m+{\cal S}} \right ]
  \frac{\nabla_j {\cal S}} {(m+{\cal S})^2}, 
\end{eqnarray}
where $i=1,2,3.$ This is the Klein-Gordon equation in hydrodynamical
form for its particle and antiparticle probability currents in external fields
for the case with the suppression of pair production.
The first two terms on the left-hand side correspond to $\partial_\mu
T_\pm^{\mu i}$ with $T_\pm^{\mu i}=\epsilon_\pm u_\pm^\mu u_\pm^i$ for
$\mu=0,1,2,3$.   The third term on the left-hand side is the quantum
stress tensor arising from the spatial variation of the amplitude of
the single-particle wave function \cite{Won76},
\begin{eqnarray}
p_{\pm\, ij}^{(q)} =-\frac{\hbar^2}{4m}\nabla^2\phi_\pm^2 \delta_{ij}
+\frac{\hbar^2}{m}\nabla_i \phi_\pm \nabla_j \phi_\pm,
\end{eqnarray}
which comes from the quantum nature of the fluid.   The first two terms on the right-hand side contain forces
coming from the scalar interaction, the electric field $F^{0i}$ and
the magnetic field $F^{ij}$, as in classical considerations.  The
third term on the right-hand side is the relativistic correction to the
time-like part of the vector interaction, and the last two terms
represent relativistic corrections associated with the spatial and
temporal variation of the scalar interaction.  Thus the dynamics of
the probability currents obey a hydrodynamical
equation, with forces on the fluid element arising from what one
expects in physical and classical considerations.  The additional element is the
presence of the quantum stress tensor $p_{ij}^{(q)}$ that comes from
the quantum nature of the fluid.

\section{Relativistic Hydrodynamics for a Many-Body System in 
the Mean-Field Description}

A many-body system in the time-dependent mean-field description
consists of a collection of independent particles moving in a
time-dependent self-consistent mean-field potential generated by all
other particles \cite{Won76,Won77,Bon74}.  Each single-particle state
$\psi_a$ is characterized by a state label $a$, particle number $\nu$, energy $E_a$, and occupation number $n_a$.  For simplicity, we consider the case
in which the mean-field potential arises from a scalar two-body interaction $v_s({\bf r}_1, {\bf r}_2)$ and a
time-like vector interaction $v_0({\bf r}_1, {\bf r}_2)$.  We  further neglect the last
three terms on the right-hand side of Eq.\ (\ref{hyd}) which represent 
higher-order relativistic corrections.  The equation of motion for 
the energy density $\epsilon_{a\nu}$ and velocity
fields $u_{a\nu}^i$ for $i=1,2,3$ in the single particle state $a$ and particle type $\nu$ is then
\begin{eqnarray}
\label{hyd24}
  & &\partial_t ( \epsilon_{a\nu} u_{a\nu}^0 u_{a\nu}^i)
  + \sum_{j=1}^3 \nabla_j  \epsilon_{a\nu} u_{a\nu}^i u_{a\nu}^j
  +\frac{m}{m+{\cal S}}   \sum_{j=1}^3 \nabla_j p_{ (a\nu)ij}^{(q)}
  \nonumber\\
  & =&
  - \phi_{a\nu}^2\nabla_i {\cal S}
  +\frac{E-e_{a\nu} A_\pm^0}{m+{\cal S}} 
  \phi_{a\nu}^2e_{a\nu} \frac{\partial A^0}{\partial x^i} ,
\end{eqnarray}
where, in the frame with fluid element at rest,
\begin{eqnarray}
{\cal S}({\bf r},t) =\int d^3{\bf r}_2~ n({\bf r_2},t) v_s({\bf r},{\bf r}_2)
\end{eqnarray}
and  
\begin{eqnarray}
A^0({\bf r},t) =\int d^3{\bf r}~ \biggl \{ n_+({\bf r_2},t) e_+
+n_-({\bf r_2},t) e_-\biggr \}
 v_0({\bf r},{\bf r}_2).
\end{eqnarray}
Here $n_+=\sum_+ n_{a+}\phi_{a+}^2$, 
$n_-=\sum_- n_{a-}\phi_{a-}^2$, and $n=n_++n_-$.
We consider a strongly interacting system in which the number of particles and antiparticles are equal so that 
 $n_+({\bf r_2})=n_-({\bf r_2})$ and  
$n_+({\bf r_2})  e_+ +n_-({\bf r_2}) e_-$ is zero.  Then the contribution from the second term on the right-hand side of Eq.\ (\ref{hyd24}) is zero.
Multiply Eq.\ (\ref{hyd24}) by $n_{a\nu}$ and sum over $\{a,\nu\}$,
we get
\begin{eqnarray}
& & \partial_t (\sum_{a\nu}n_{a\nu}\epsilon_{a\nu} u_{a\nu}^0 u_{a\nu}^i )
  + \sum_{j=1}^3 \nabla_j   (\sum_{a\nu}n_{a\nu}\epsilon_{a\nu} u_{a\nu}^i u_{a\nu}^j)
\nonumber\\
& &
  +\frac{m}{m+{\cal S}}   \sum_{j=1}^3  \nabla_j ( \sum_{a\nu}n_{a\nu}p_{(a\nu) ij}^{(q)})
  + (\sum_{a\nu}n_{a\nu}\phi_{a\nu}^2) \nabla_i {\cal S} 
=0.
\end{eqnarray}
We define the total energy density $\epsilon$ by
\begin{eqnarray}
\sum_{a\nu} n_{a\nu} \epsilon_{a\nu}=\epsilon,
\end{eqnarray}
and the average  4-velocity $u$ by 
\begin{eqnarray}
u = {\sum_{a\nu}  n_{a\nu} \epsilon_{a\nu} u_{a\nu}}
       /{\epsilon}.
\end{eqnarray}
We can introduce the thermal stress tensor $p_{ij}^{(t)}$ for $\{i,j\}=1,2,3$ as the correlation of the  deviations of the single-particle velocity fields from the average 
\begin{eqnarray}
\sum_{a\nu} n_{a\nu} \epsilon_{a\nu} (u_{a\nu}^i-u^i) (u_{a\nu}^j -u^j)
\equiv p_{ij}^{(t)}.
\end{eqnarray}
For the case with the suppression of pair production, we obtained the equation of hydrodynamics 
\begin{eqnarray}
\partial_t (\epsilon u^0 u^i )
+ \sum_{j=1}^3 \left \{ \nabla_j \left  (\epsilon u^i u^j +p_{ij}^{(t)}+p_{ij}^{(v)}\right )
+\frac{m}{m+{\cal S}}  \nabla_j  p_{ ij}^{(q)} \right \}
=0 ,
\end{eqnarray}
where the total quantum stress tensor is 
\begin{eqnarray}
p_{ ij}^{(q)} =-\frac{\hbar^2}{4m}\nabla^2\sum_{a\nu} n_{av} \phi_{a\nu}^2 \delta_{ij}
+\frac{\hbar^2}{m}\sum_{a\nu} n_{av} \nabla_i \phi_{a\nu} \nabla_j \phi_{a\nu},
\end{eqnarray}
and the pressure due to the interaction $p_{ij}^{(v)}$ is 
\begin{eqnarray}
\frac{\partial}{\partial x^j} p_{ij}^{(v)}({\bf r},t)=n({\bf r},t)\nabla_i {\cal S}({\bf r},t)=
n({\bf r},t) \frac{\partial}{\partial x^j}
\int d^3 {\bf r}_2 n({\bf r}_2,t) v_s({\bf r},{\bf r}_2).
\end{eqnarray}
The mean-field stress tensor $p_{ij}^{(v)}$ can also be given as
\begin{eqnarray}
 p_{ij}^{(v)}=
\left \{ n \frac{\partial (W^{(v)} n) }{\partial n}
- W^{(v)} n \right \}\delta_{ij}
\end{eqnarray}
where $W^{(v)}$ is the energy per particle arising from the mean-field
interaction.  As an illustrative example, we can consider a
density-dependent two-body interaction
\begin{eqnarray} 
v_s( {\bf r},{\bf r}_2)=[a_2+a_3 n(({\bf r}+{\bf r}_2)/2)]
\delta({\bf r}-{\bf r}_2).
\end{eqnarray}
The contribution of the mean-field interaction to the stress tensor is then
\begin{eqnarray}
 p_{ij}^{(v)}=
 \frac{1}{2}[a_2+2a_3 n({\bf r})] {n^2} \delta_{ij},
\end{eqnarray}
whose magnitude increases with the density and the strengths of the interaction. 

The thermal stress tensor 
 $p_{ij}^{(t)}$ can take on different values,
depending on the 
occupation numbers of the single-particle states that 
determine the degree of thermal equilibrium of the system.
In the time-dependent mean-field description, the motion of
each particle state can be individually followed \cite{Bon74}. 
The
occupation numbers $n_{a\nu}$  of the single-particle states will remain
unchanged, if there are no additional interaction between the single
particles due to the residual interactions.  When particle residual
interactions are allowed as in the extended time-dependent mean-field
approximation \cite{Won78}, the occupation numbers will change and
will approach an equilibrium distribution as time proceeds.

We note that as in the non-relativistic case, the total pressure arises from many sources. We come again to the observation that in situations when $|p_{ij}^{(q)}+p_{ij}^{(v)}| \gg p_{ij}^{(t)}$ for a strongly-coupled system at low and moderate temperatures, 
 there can be situations when the  system behaves quasi-hydrodynamically, even though the state of the system has not yet reached thermal equilibrium.  In this case, the hydrodynamical state is maintained mainly by the quantum stress tensor and the strong mean fields.

\section{Summary and Discussions}

For dense systems with strongly interacting constituents, a reasonable
description of the system can be formulated in terms of constituents
moving in the strong mean fields generated by all other particles.
From such an analysis, we find that the probability currents of the
system obey a hydrodynamical equation with the stress tensor arising
from many contributions.  There is the quantum stress tensor that
arises from quantum effects, there is the thermal stress tensor that
arises from the deviation of the single-particle velocity fields from
the average velocity fields, and there is the mean-field stress tensor that
arises from the mean-field interactions.

The importance of the three different contributions depend on the
physical situations that are present in the system.  For low
temperature dynamics for which the quantum effects and mean fields are
important, the dynamics of the strongly-coupled system will be
dominated by the quantum stress tensor and the mean-field stress
tensor.  In that case, the degree of thermalization is of less
significance as the thermal stress tensor plays a minor role.  On the
other hand, for very high temperatures for which the magnitude of the
temperature far exceeds the strengths of the mean-field interactions
and the quantum pressure, the thermal stress tensor plays the dominant
role.  The magnitude of the thermal stress tensor will depend
sensitively on the degree of thermalization of the system.  In between
these limits, one can envisage the transition from the quantum and
mean field dominating strongly-coupled regime to the thermal pressure
dominating  weakly-coupled regime as temperature increases.

What we have discussed is only a theoretical  framework that exhibits clearly the different sources of stress tensors.  To study specifically the dynamics of the quark-gluon plasma for example, it will be necessary to investigation the specific nature of different constituents and their interactions.  Nevertheless, the general roles played by the different components of stress tensors can still be a useful reminder on the importance of the quantum and mean-field stress tensors
in the strongly-coupled regime, at temperature just above the transition temperature $T_c$.  The dominance of the quantum and mean-field stress tensors over the thermal stress tensor may imply that hydrodynamics may commence at an early stage even when thermal equilibrium has not been reached, as the total stress tensor may be hardly affected by the variation of the thermal stress tensor.   It will therefore be of great interest to study whether the fast onset of the hydrodynamical motion as suggested by the elliptic flow measurements \cite{Oll08} may be caused by the dominance of the mean-field and quantum stress tensors for a strongly-coupled quark-gluon plasma.  For $T$ slightly greater than $T_c$ such a possibility may perhaps be the case.

%%%%%%%%%%%%%%%%%%%%%%%%%%%%%%%%%%%%%%%%%%%%%%%%
%% BACKMATTER
%%%%%%%%%%%%%%%%%%%%%%%%%%%%%%%%%%%%%%%%%%%%%%%%
\vspace*{-0.3cm}
\begin{theacknowledgments}
The author wishes to thank Prof. H. W. Crater  and for helpful discussions.  
\end{theacknowledgments}



\bibliographystyle{aipproc}   % if natbib is available

%\bibliography{sample}

%  \typeout{******************************************}
%  \typeout{** Please run "bibtex \jobname" to optain}
%  \typeout{** the bibliography and then re-run LaTeX}
%  \typeout{** twice to fix the references!}
%  \typeout{******************************************}

%\end{document}

%%%%%%%%%%%%%%%%%%%%%%%%%%%%%%%%%%%%%%%%%%%
%% The following lines show an example how to produce a bibliography
%% without the help of the BibTeX program. This could be used instead
%% of the above.
%%%%%%%%%%%%%%%%%%%%%%%%%%%%%%%%%%%%%%%%%%%

\end{document}